\documentclass[astrosymb, preprint]{aastex7} %add linenumbers for submission

\usepackage{enumitem} % Lists
    \setlist[enumerate]{noitemsep} %list separation

\begin{document}

\title{The Relationship Between Eddington Ratio and Column Density in U/LIRG AGN}

\author[0009-0002-6248-3688]{Jaya Nagarajan-Swenson}
\affiliation{University of Virginia, 530 McCormick Road, Charlottesville, VA 22903, USA}
\email{jaya@virginia.edu}

\author[0000-0003-3474-1125]{George C. Privon}
\affiliation{National Radio Astronomy Observatory, 520 Edgemont Road, Charlottesville, VA 22903, USA}
\affiliation{Department of Astronomy, University of Florida, 1772 Stadium Road, Gainesville, FL 32611, USA}
\affiliation{University of Virginia, 530 McCormick Road, Charlottesville, VA 22903, USA}
\email{gprivon@nrao.edu}

\author[0000-0003-2638-1334]{Aaron S. Evans}
\affiliation{University of Virginia, 530 McCormick Road, Charlottesville, VA 22903, USA}
\affiliation{National Radio Astronomy Observatory, 520 Edgemont Road, Charlottesville, VA 22903, USA}
\email{ae3f@virginia.edu}

\author[0000-0003-0057-8892]{Loreto Barcos-Mu\~{n}oz}
\affiliation{National Radio Astronomy Observatory, 520 Edgemont Road, Charlottesville, VA 22903, USA}
\affiliation{University of Virginia, 530 McCormick Road, Charlottesville, VA 22903, USA}
\email{lbarcos@nrao.edu}

\author[0000-0001-5231-2645]{Claudio Ricci}
\affiliation{Instituto de Estudios Astrof\'isicos, Facultad de Ingenier\'ia y Ciencias, Universidad Diego Portales, Av. Ej\'ercito Libertador 441, Santiago, Chile}
\affiliation{Kavli Institute for Astronomy and Astrophysics, Peking University, Beijing 100871, China}
\email{claudio.ricci@mail.udp.cl}

\author[0000-0001-7421-2944]{Anne M. Medling}
\affiliation{Department of Physics \& Astronomy and Ritter Astrophysical Research Center, University of Toledo, Toledo, OH 43606, USA}
\email{anne.medling@utoledo.edu}

\author[0000-0002-1912-0024]{Vivian U}
\affiliation{IPAC, Caltech, 1200 E. California Blvd., Pasadena, CA 91125, USA}
\affiliation{Department of Physics and Astronomy, University of California, 4129 Frederick Reines Hall, Irvine, CA 92697, USA}
\email{vivianu@uci.edu}

\author[0000-0003-4546-3810]{Alejandro Saravia}
\affiliation{University of Virginia, 530 McCormick Road, Charlottesville, VA 22903, USA}
\email{mas6um@virginia.edu}

\author[0009-0002-2049-9470]{Kara N. Green}
\affiliation{University of Virginia, 530 McCormick Road, Charlottesville, VA 22903, USA}
\email{zas6xr@virginia.edu}

\author[0000-0001-7690-3976]{Makoto Johnstone}
\affiliation{University of Virginia, 530 McCormick Road, Charlottesville, VA 22903, USA}
\email{fhh3kp@virginia.edu}

\author[0009-0006-6594-1516]{Gabriela A. Meza}
\affiliation{Universidad Nacional Aut\'onoma de Honduras, Ciudad Universitaria, Tegucigalpa, Honduras}
\email{gabrielamezar02@gmail.com}

\begin{abstract}
The local X-ray AGN population appears to follow a growth cycle regulated by the AGN's own radiation, marked by changes in their obscuration and Eddington ratio during accretion events. Because AGN in infrared-selected galaxies are more likely to be Compton-thick and have evidence for over-massive black holes, we explore whether infrared-selected AGN follow the radiation-regulated AGN growth scheme. We calculate the Eddington ratios of nine U/LIRG AGN with dynamical BH mass measurements, finding that though the number of objects is limited, AGN in IR-selected galaxies appear consistent with radiation pressure-regulated growth. We suggest that enlarging the sample of dynamical BH mass measurements in IR-selected systems will provide more stringent tests of whether their AGN are primarily regulated by radiation pressure.
\end{abstract}

\section{Introduction} \label{sec:intro}  
\begin{figure*}[ht]
    \centering
    \includegraphics[width=0.9\linewidth]{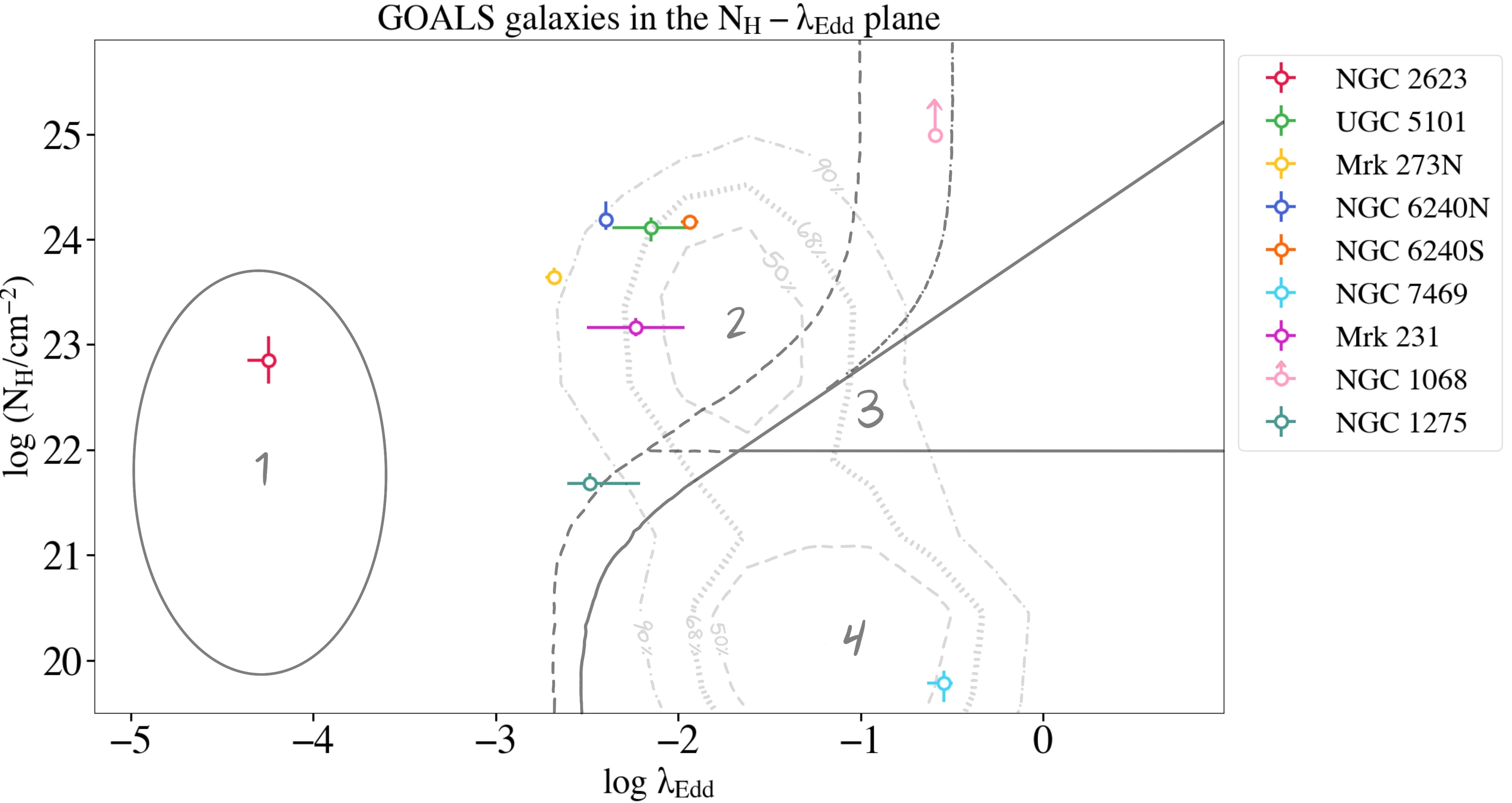}
    \caption{GOALS systems in the $N_H - \lambda_{Edd}$ plane. Colored symbols show U/LIRG AGN, and light grey contours show BASS AGN \citep{Ricci2022}. Sources cycle clockwise in this plane, from (1) an accretion event to (2) an obscured phase, (3) a blowout phase, and (4) an unobscured phase. Dark grey lines delimiting the blowout region from the obscured phase show hypothetical Eddington limits for dusty gas \citep[see][for details]{Ricci2021}, and the horizontal line at $N_H=22$ represents the maximum possible contribution of a host galaxy to $N_H$.}
    \label{fig:eddingtonrat}
\end{figure*}

Observations of Active Galactic Nuclei (AGN) in \cite{Ricci2017} show that radiation from an AGN plays a key role in determining and regulating its obscuration. \cite{Fabian2008} suggested that radiation pressure drives this relationship, and that feedback from the AGN is strengthened by dust mixed with the surrounding gas. \cite{Ricci2022} demonstrated that X-ray selected AGN from the BASS sample \citep{Ricci2017, Koss2022} show a reduction in the fraction of obscured sources at high Eddington ratio, consistent with radiation pressure on dusty gas being a key feedback mechanism. These works established the evolutionary cycle illustrated in Figure \ref{fig:eddingtonrat}, wherein AGN cycle through phases of high and low obscuration and Eddington ratio $\lambda_{Edd}$ during a growth episode. In this cycle, 
\begin{enumerate}
    \item An accretion event funnels material towards an AGN.
    \item Some of the material obscures the AGN, increasing its column density $N_{H}$, and some material is accreted, increasing $\lambda_{Edd}$. 
    \item When the AGN reaches the Eddington limit for dusty gas \citep{Fabian2009, Ishibashi2018}, radiation pressure expels the surrounding material, leading to a brief period of high accretion rate at decreasing column density. 
    \item This blowout leaves the AGN unobscured but still growing.
\end{enumerate}
Once all available material has been accreted or expelled, AGN growth ceases. This framework requires obscuring material to be within the sphere of influence of the supermassive black hole \citep[SMBH; within 65 pc of a $10^{9}$ M$_{\odot}$ SMBH,][]{Ricci2017}, allowing it to be cleared away by the high $\lambda_{Edd}$. 

In principle, the processes regulating AGN growth happen within an SMBH's sphere of influence so any large-scale host galaxy activity should have minimal impact. However, \cite{Ricci2022} suggests that AGN growth in (Ultra)Luminous InfraRed Galaxies (U/LIRGs, with $L_{IR}>10^{11}L_{\odot}$ and $10^{12}L_{\odot}$) may follow a different pathway. Most U/LIRGs are galaxy mergers \citep[e.g.,][]{Armus1987,Armus2009}, which funnel gas towards the center of a galaxy, stimulating AGN activity and resulting in higher typical obscuration than X-ray AGN \citep{Ricci2021}. This high column density is interpreted as the obscuring material in U/LIRGs residing at larger scales than in X-ray AGN. 
Moreover, U/LIRG SMBHs are overmassive relative to standard scaling relations \citep{Medling2015}, implying accelerated SMBH growth during a merger or the presence of massive gas reservoirs within the SMBH sphere of influence \citep{Medling2019}.
These factors combine to imply that the AGN fueling and feedback cycle in U/LIRG systems may differ from those in the X-ray AGN of \cite{Ricci2022}. In this note, we present $\lambda_{Edd}$ for nine U/LIRG AGN from the Great Observatories All-sky LIRG Survey \citep[GOALS;][]{Armus2009} to determine whether infrared-selected AGN follow the same radiation-regulated growth cycle as X-ray AGN.

%%%%%%%%%%%%%%%%%%%%%%%%%%%%%%%%%%%%%%%
%%%%%%%%%%%%%%%%%%%%%%%%%%%%%%%%%%%%%%%
\section{Black Hole Mass and Eddington Ratio Measurements}
\label{sec:analysis}

The focus of this analysis, $\lambda_{Edd}$, requires robust measurements of an AGN's mass and luminosity. Because U/LIRG AGN are often heavily obscured, high-energy observations are most effective at determining their luminosity. Thus, the sample in this study is comprised of the nine U/LIRG AGN with luminosities and column densities measured at 10-24 keV \citep{Ricci2021} and directly measured dynamical masses. 

Five SMBHs were weighed using dynamical gas measurements in the central $\sim$25 pc of each galaxy in \cite{Medling2015}. These masses likely include some gas; however, analysis of NGC$\,$6240 in \cite{Medling2019} determined the gas contribution to its measured dynamical mass is minimal. The remaining four were weighed using reverberation mapping \citep[NGC$\,$7469,][]{Lu2021}, broad-line region polarization \citep[Mrk$\,$0231,][]{Afanasiev2019}, accretion disk water masers \citep[NGC$\,$1068,][]{Lodato2003}, and molecular hydrogen kinematics \citep[NGC$\,$1275,][]{Scharwachter2013}. 

Using these data, we measure $\lambda_{Edd}$ in nine U/LIRGs and place them into the radiation-regulated growth framework from \cite{Ricci2022} in Figure \ref{fig:eddingtonrat}. Of this sample, three are ULIRGs and five are LIRGs \citep{DiazSantos2017}. NGC$\,$7469 is an early-stage merger, NGC$\,$1068 and NGC$\,$1275 are non-merger LIRGs, and the remainder are late-stage mergers \citep[as determined from the \textit{H}-band,][]{Haan2011}.

%%%%%%%%%%%%%%%%%%%%%%%%%%%%%%%%%%%%%%%
%%%%%%%%%%%%%%%%%%%%%%%%%%%%%%%%%%%%%%%
\section{Discussion}
\label{sec:discussion}
All U/LIRG AGN fall within the same $\lambda_{Edd}$ and $N_H$ ranges as X-ray AGN, suggesting that U/LIRG AGN are consistent with their accretion being regulated by radiation pressure. 
Additionally, we find no clear correlation between merger stage and an AGN's position within the radiation-regulated feedback cycle. However, due to the small sample, we cannot draw firm conclusions about the relationship between AGN accretion phase and merger progression. An expanded sample would allow a robust analysis of these trends, offering insight into the timescales of accretion and obscuration during galaxy mergers. 

Excluding NGC 7469, U/LIRG systems lie within a relatively high-obscuration regime of their respective phases in the $N_H - \lambda_{Edd}$ plane. While this indicates that U/LIRG AGN are more obscured than their X-ray counterparts, that is a likely result of GOALS' infrared selection \citep{Ricci2021}. To become a U/LIRG, there must be enough material re-emitting in the infrared to drive $L_{IR}$ above $10^{11}$ L$_{\odot}$, thus these objects are necessarily highly obscured. 

Overall, AGN from the IR-selected GOALS sample fall in regions of this radiation-regulated growth framework similar to those of the X-ray selected BASS survey. We conclude that despite their potentially overmassive SMBHs \citep{Medling2015} and high column densities \citep{Ricci2021}, U/LIRG AGN remain consistent with the radiation-regulated growth framework presented for X-ray AGN in \cite{Ricci2022}. 

\begin{acknowledgements}
    J. N. S. acknowledges support from NASA NuSTAR awards 80NSSC21K1177 and 80NSSC22K0064, and JWST-GO-03368.008-A. This research has made use of the NASA/IPAC Extragalactic Database (NED) and NASA’s Astrophysics Data System.
\end{acknowledgements}

\bibliography{bib}{}
\bibliographystyle{aasjournalv7}

\end{document}